*Research article*

# A novel semi-supervised multi-view clustering framework for screening Parkinson's disease

**Xiaobo Zhang [1,2], Donghai Zhai [1,2,\*], Yan Yang [1,2], Yiling Zhang [1,2] and Chunlin Wang [1,2]**

1   School of Information Science and Technology, Southwest Jiaotong University, Chengdu 611756, China

2   National Engineering Laboratory of Integrated Transportation Big Data Application Technology, Southwest Jiaotong University, Chengdu 611756, China

\*   **Correspondence:** Email: dhzhai@swjtu.edu.cn; Tel: +86-28-66366200.

**Abstract:** In recent years, there are many research cases for the diagnosis of Parkinson's disease (PD) with the brain magnetic resonance imaging (MRI) by utilizing the traditional unsupervised machine learning methods and the supervised deep learning models. However, unsupervised learning methods are not good at extracting accurate features among MRIs and it is difficult to collect enough data in the field of PD to satisfy the need of training deep learning models. Moreover, most of the existing studies are based on single-view MRI data, of which data characteristics are not sufficient enough. In this paper, therefore, in order to tackle the drawbacks mentioned above, we propose a novel semi-supervised learning framework called Semi-supervised Multi-view learning Clustering architecture technology (SMC). The model firstly introduces the sliding window method to grasp different features, and then uses the dimensionality reduction algorithms of Linear Discriminant Analysis (LDA) to process the data with different features. Finally, the traditional single-view clustering and multi-view clustering methods are employed on multiple feature views to obtain the results. Experiments show that our proposed method is superior to the state-of-art unsupervised learning models on the clustering effect. As a result, it may be noted that, our work could contribute to improving the effectiveness of identifying PD by previous labeled and subsequent unlabeled medical MRI data in the realistic medical environment.

## 1. Introduction

Parkinson's disease (PD) is a degenerative and disabling disease in the nervous system [1], which generally occurs in the elderly. Its clinical manifestations mainly include quiescent tremor, motor retardation, myotonia and postural gait disorder. PD not only affects the life quality of a patient, but also brings heavy burden to his/her family and the society. According to medical statistics, PD is rare in young populations aged under 40 and the average age among people with PD is 60 years old, of which incidence rate tends to increase with age [2] – [5].

For the elderly, screening PD as early as possible is very vital for prevention and delaying progress to assist in auxiliary diagnosis. At present, the diagnosis of PD mainly relies on the clinical symptoms of patients and the professional knowledge of clinical neurologists. However, some missed diagnosis and misdiagnosis may happen due to the complexity of pathology in PD [6]. Most doctors would recommend inspecting a neuroimaging examination before the formal clinical diagnosis of PD, containing magnetic resonance imaging (MRI), functional magnetic resonance imaging (fMRI), and positron emission tomography (PET), etc. In this study, we evaluate the proposed method on the MRI data obtained from the Parkinson's Progression Markers Initiative (PPMI) platform [7].

Nowadays, there are some machine learning techniques which are used to automatically diagnose PD and predict clinical diagnostic scores. Shi et al. constructed a novel cascaded multicolumn RVFL+ (cmcRVFL+) framework for the single-modal neuroimaging-based diagnosis of PD to reduce the difficulty of data acquisition [8]. Peng et al. [9] used a multilevel-ROI-features-based machine learning method to detect sensitive morphometric biomarkers in PD. Prashanth et al. [10] developed a classification model based on machine learning techniques to partition degenerative early PD class and healthy normal/non-degenerative condition class. Oliveira et al. [11] studied a fully automatic computational solution for computer-aided diagnosis of PD with the technique of support vector machines and a voxel-as-feature approach based on single photonemission computed tomography brain images. Garraux et al. [12] used relevance vector machine in combination with booststrap resampling for nonhierarchical multiclass classification based on fluorodeoxyglucose positron emission tomography scans performed in patients of PD. Long et al. [13] proposed a non-invasive technology intended for using in the diagnosis of early PD by integrating the advantages of various models with multi-modal MRI data. Abs et al. [14] investigated connection-wise patterns of functional connectivity to distinguish Parkinson's disease patients according to their cognitive status by using machine learning methods. Lei et al. [15] discussed a joint regression and classification framework for PD diagnosis via magnetic resonance and diffusion tensor imaging data. They devised a unified multi-task feature selection model to explore multiple relationships among features, samples, and clinical scores. Adeli et al. [16] proposed an approach to diagnose PD with MRI data using a feature-sample selection (JFSS) method and a robust classification framework. Adeli et al. [17] also investigated a joint kernel-based feature selection and a classification framework for early Diagnosis of PD on multi-modal neuroimaging data.

However, the existing approaches on automatic diagnosis mainly focus on classification or prediction of PD. The study of auxiliary screening and the diagnosis of PD on MRI data with semi-supervised multi-view clustering approaches is scarce, and there are few screening model working on preventing and delaying the PD deterioration. In this work, we propose a novel semi-supervised multi-view learning framework with clustering based on the Robust Multi-View K-Means Clustering (RMKMC) [18] by the cross-validation method [19], namely **S**MC (as shorthand for **S**emi-supervised **M**ulti-view **C**lustering architecture). We train and test our model using multi-view brain MRI data information after data representation with the Linear Discriminant Analysis technology (known as LDA) [20] and the sliding window method [21]. Compared with the existing clustering methods, such as Gaussian Mixture Model (GMM) [22], K-Means [23], K-Medoids [24], Agglomerative Clustering algorithm (AC) [25], Balanced Iterative Reducing and Clustering Using Hierarchies (Birch) [26] and Spectral Clustering algorithm (SC) [27], our model achieves the best results and effectively partitions three types of samples (i.e., non-patients, prodromal PD and confirmed PD). The results are useful to help doctors for auxiliary diagnosis, especially in screening latent PD, which can promote early diagnosis and treatment, delay disease progression and further reduce the occurrence of Parkinson's disease.

In summary, this paper makes the following contributions:

1) It proposes a novel semi-supervised multi-view learning architecture with clustering (SMC) for screening Parkinson's disease.

2) The proposed SMC uses the dimensionality reduction algorithm of Linear Discriminant Analysis (LDA) and the sliding window method to extract effective feature information from the MRI data. In such a way, this supervised method could make full use of the color and texture features in MRI images and avoid overfitting issues caused by data dimension imbalance.

3) It exploits multi-view learning technology on the proposed semi-supervised learning architecture. Multi-view learning is not only able to capture key features from each single view, but also can integrate the comprehensiveness of MRI features.

The rest of this paper is organized as follows. Section 2 introduces the preprocessed dataset, our proposed SMC model and experimental design. Section 3 presents the extensive experiment results. Furthermore, the discussion is conducted in Section 4. Finally, Section 5 concludes this paper.

## 2. Materials and Method

### 2.1. MRI Datasets and Data Preprocessing

2.1.1. MRI Datasets

In this paper, the MRI datasets from the PPMI database are adopted for experiments. For the latest and more data information on research, please visit www.ppmiinfo.org [7]. For the purpose of this study, all cross-sectional MRI datasets are screened. Our MRI data contains 3 classes, which includes 204 healthy normal samples, 193 prodromal PD samples and 201 confirmed PD samples, respectively. Because of the swallowtail sign that provides auxiliary basis in PD diagnosis [28], the samples reflecting the swallowtail area are chosen from each subject. And each sample of all the three subjects is preprocessed with relevant dimensional reduction technologies in data processing. A few initial data of three subjects are presented in Figure 1 for more intuitive and effective description.

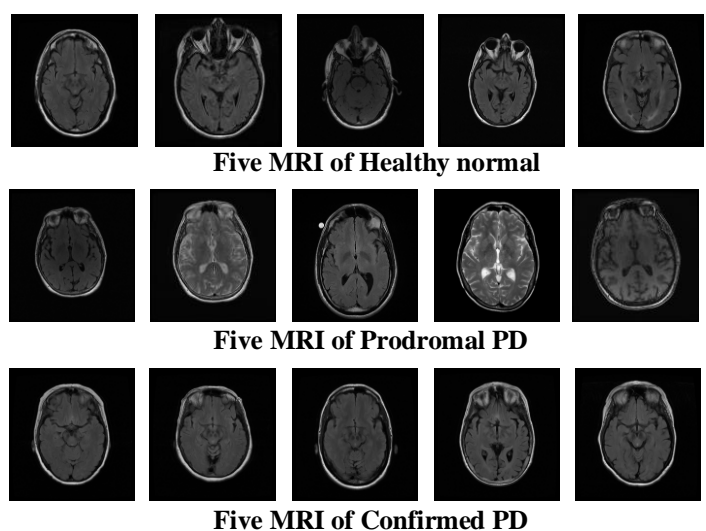

**Five MRI of Healthy normal**

**Five MRI of Prodromal PD**

**Five MRI of Confirmed PD**

**Figure 1. Five MRI of three subjects.**

2.1.2.  Data Preprocessing

In order to improve the accuracy and performance of our method in experiments, the original MRI data sets need to be standardized. The initial datasets we selected are preprocessed by a few methods. The specific processing steps mainly include: obtaining gray value image, using median filter to remove noise points, acquiring linear normalization gray value, and eventual extracting Region Of Interest (ROI) shown in Figure 2 by detecting the edge features of brain MRI. Also, the swallow tail area for each sample is signed. More obviously, the swallow tail areas of two MRI for the three classes are showed in Figure 3.

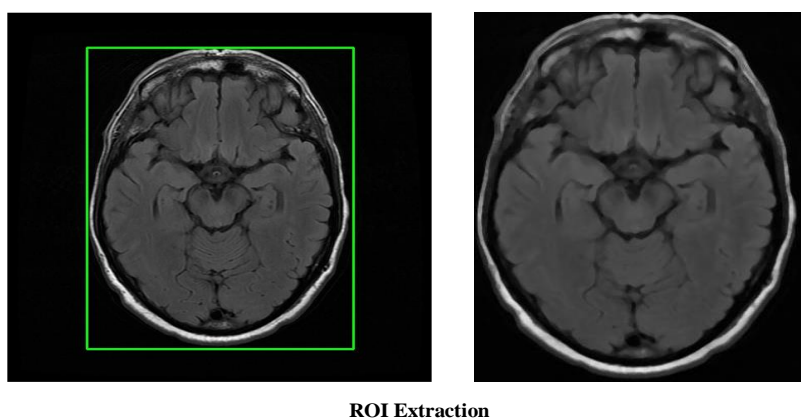

**ROI Extraction**

**Figure 2. ROI Extraction for one MRI sample.**

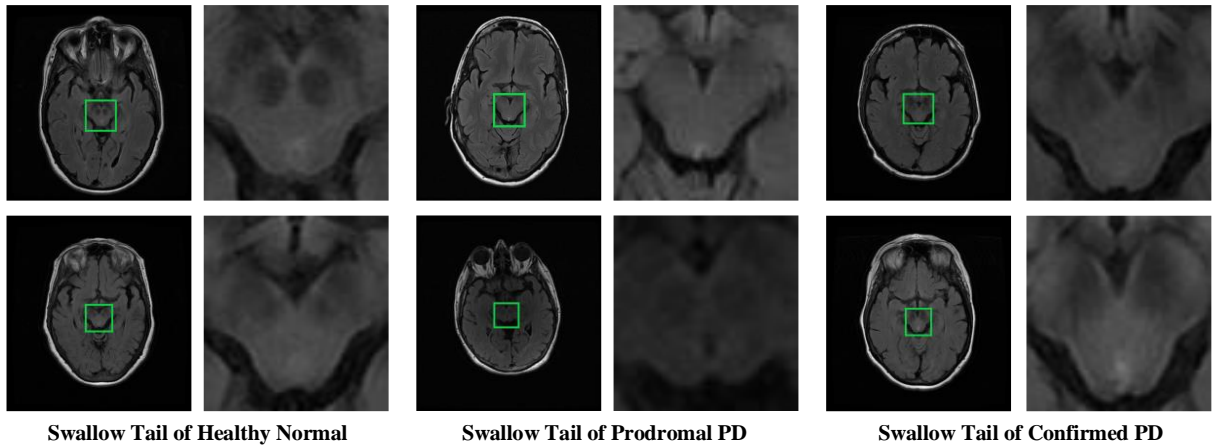

**Swallow Tail of Healthy Normal**  **Swallow Tail of Prodromal PD**  **Swallow Tail of Confirmed PD**

**Figure 3. The swallow tail areas of two MRI for the three subjects.**

*2.2. Feature Extraction*

In order to make full use of the color and texture features in the MRI data sets, the 7x7 sliding window method is utilized for feature extraction, as the Figure 4 shown. The initial position of the green box is the upper left corner of ROI, after the sub image is extracted, the window is moved from left to right, from top to bottom, until the lower right corner of ROI is overlapped with the red box [21].

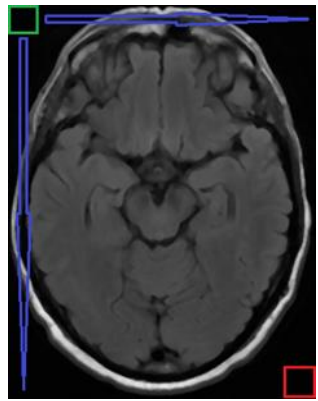

**Figure 4. The process of extracting feature using 7x7 sliding window method.**

Each subimage is quantized with 16 level gray scale. After that, the 4 different types of texture features in gray level co-occurrence matrix are obtained, including contrast, homogeneity, energy and correlation [29]. And the 3 categories of statistical information regarding to standard deviation i.e. sigma, skew and kurtosis are also extracted. As a result, 7 different types of features of the MRIs are extracted from each sliding window. Simultaneously, we reconstruct 7 types of features as 7 data views, of which dimension size of each view is equal to the number of image pixels and each feature is independent with each other. Finally, seven views representing different categories of features are shown in the Figure 5.

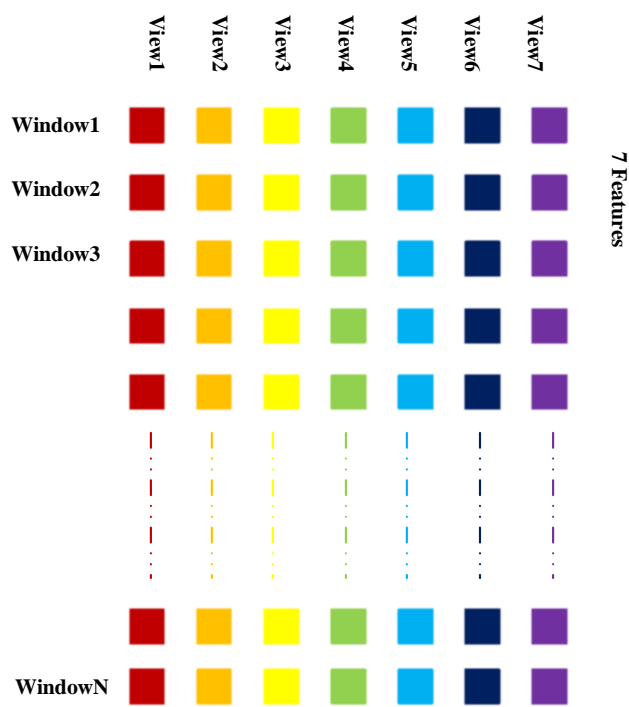

**Figure 5. The seven views extracted from one MRI.**

*2.3. Our proposed SMC*

In this study, we mainly integrate 3 machine learning techniques into our SMC framework to assist in auxiliary diagnosis of PD, including the Cross-Validation technology, the dimensionality reduction technology and the robust multi-view K-Means clustering technology. In this section, we will review these methods and give the detailed framework of our proposed SMC.

2.3.1. Review relevant methods

**The Cross-Validation technology.** Cross-Validation is a method widely used in machine learning to build models and verify model parameters [5]. It is a statistical method to evaluate and compare the algorithms by dividing data into two parts: one for training model, and the other for validating model. In a typical Cross Validation, training and validation sets must cross in successive rounds so that every data point can be validated. The basic form of Cross Validation is K-fold Cross Validation. Other forms are special cases of K-fold Cross Validation, or repeated rounds related to it [30].

**The Dimensionality Reduction technology.** There are two main methods of dimensionality reduction: unsupervised method and supervised method. For an unsupervised method, the labels of data could not be tagged, which means that we can only classify or cluster the data samples by learning similar features among samples, while for a supervised approach, class labels are considered [31] to obtain more robust classification or clustering results. There are many unsupervised dimensionality reduction technologies, such as independent component analysis (ICA) [32] and non-negative matrix factorization (NMF) [33], but the most dominantone is principal component analysis (PCA) [34]. PCA can reduce the dimension of data while retaining most of the changes in the data set. In addition, many supervised dimensionality reduction technologies are proposed, such as mixed discriminant analysis (MDA) [35] and neural network (NN) [36], but the most famous one is linear discriminant analysis (LDA) [37]. LDA can improve the calculation efficiency in the process of data analysis and reduce the overfitting caused by the increase of dimensionality.

**The Robust Multi-View K-Means Clustering technology.** There are many multi-view clustering methods [38] to utilize the features from multiple views and enhance the experimental performance. The Robust Multi-View K-Means Clustering (RMKMC) we selected in this paper is based on the K-Means clustering method with robust and multi-view knowledge. RMKMC could integrate the heterogeneous features for clustering and solve the large-scale multi-view clustering problem. Utilizing the common cluster indicator, RMKMC could search a consensus pattern and do clustering across multiple visual feature views. Moreover, this method is robust to the outliers in input data, and learns the weights of each view adaptively [18].

### 2.3.2. Semi-supervised Multi-view learning architecture with Clustering

In this paper, we propose a novel Semi-supervised Multi-view learning framework with Clustering (SMC) based on RMKMC [18] which avoids sampling bias and achieves better the effect of dimensionality reduction and superior performance of semi-supervised clustering. The proposed SMC exploit cross-validation strategy to divide the multi-view samples into training data with labels and testing data unlabeled. The training data is used to train the supervised dimensionality reduction model and then the testing data is inputted into the model to obtain the reduction results. Finally, a clustering method is utilized to partition the test data into several clusters. The SMC is not able to extract the effective features among MRIs, but also avoids to collect massive data samples in the field of PD. And more detailed description about the algorithm framework of SMC is shown in table 1.

**Table 1. The algorithm framework of SMC.**

**Algorithm 1 The algorithm of SMC:**

**Input:**
  1. Data for $M$ views $\{X^{(1)}, \cdots, X^{(v)}\}$ and $X^{(v)} \in \mathbb{R}^{d_v \times n}$.
  2. Label of each trained sample.

**Output:**
  1. Reduced dimension data of labeled dataset for $M$ views $\{W^{(1)}, \cdots W^{(v)}\}$.
  2. The clustering result for each view.

**Steps:**

  For each $X^{(v)}$ in $\{X^{(1)}, \cdots, X^{(v)}\}$

  1. Divide the dataset $X^{(v)}$ into labeled dataset $X'^{(l)}$ and unlabeled dataset $X'^{(u)}$.
  2. Use data and labels of $X'^{(l)}$ to train LDA model.
  3. Dimension reduction of $X'^{(u)}$ using trained LDA model, Add results to $W^{(v)}$.
  4. Clustering with data $W^{(v)}$.

*2.4. Experimental design*

In the experiment, we compare our SMC algorithm with several classical single-view clustering methods including GMM, K-Means, K-Medoids, AC, Birch and SC to verify the effect of our multi-view technique. And the unsupervised dimensionality reduction method like PCA is utilized in our experiments to compare with LDA and demonstrate the performance of semi-supervised learning in screening PD. In addition, the 5-fold cross validation strategy is selected to train semi-supervised model [5].

Specifically, to demonstrate that the proposed semi-supervised framework can mine more accurate information and are superior to unsupervised methods, the PCA which is an unsupervised technique is selected to reduce the dimension of the extracted features [34] to compare with the SMC framework. Meanwhile, the number of training dataset with label and testing dataset without label is divided into 80% and 20%. The results of the two groups of experiments are compared and analyzed with different clustering methods and data views, separately. Besides, the experimental results of each clustering and data view are compared only by SMC framework to show the performance of our model. Figure 6 gives the more detailed process of the experimental design.

There are a lot of clustering result evaluation indexes for clustering experiment, but the factors of the requirement of medical disease prediction need to be considered. Three standard clustering evaluation metrics are chosen to measure the clustering performance, that is, Clustering Accuracy (Acc), Fowlkes-Mallows score (FM) and Adjusted Rand index (Rand) [39].

Acc is the proximity value of the clustering results, which could evaluate the accuracy of the cluster. The Acc is defined as follows:

$$Acc = \sum_{k=1}^{K} \frac{N_k}{N} \quad (0 \leq MP \leq 1) \tag{1}$$

Where $N_k$ is the number of data items that are correctly classified to each class. The bigger the value of MP is, the better the clustering performance is.

The FM is defined as the geometric mean of the pairwise precision and recall:

$$FMI = \frac{TP}{\sqrt{(TP + FP)(TP + FN)}} \tag{2}$$

where $TP, FP, FN$ are true positive, false positive and false negative, respectively.

And the Rand index needs to give the actual category information $C$, assuming $K$ is the clustering result, $a$ represents the logarithms of elements of the same category in both $C$ and $K$, and $b$ represents the logarithms of elements of different categories in $C$ and $K$. The Rand index is below:

$$RI = \frac{a + b}{C_2^{n_{samples}}} \tag{3}$$

Among them, the total number of element pairs that can be composed in the $C_2^{n_{samples}}$ data set. The value range of $RI$ is $[0,1]$. For random results, the Rand index cannot guarantee that the score is close to zero. In order to achieve the goal that when the clustering results are randomly generated, the index should be close to zero, Adjusted Rand index (Rand) is proposed, which has a higher degree of discrimination. ARI is shown as follows:

$$ARI = \frac{RI - E(RI)}{max(RI) - E(RI)} \tag{4}$$

The value range of $ARI$ is $[-1,1]$. A larger value means that the clustering result is more consistent with the real situation. In a broad sense, $ARI$ measures how well two data distributions fit.

9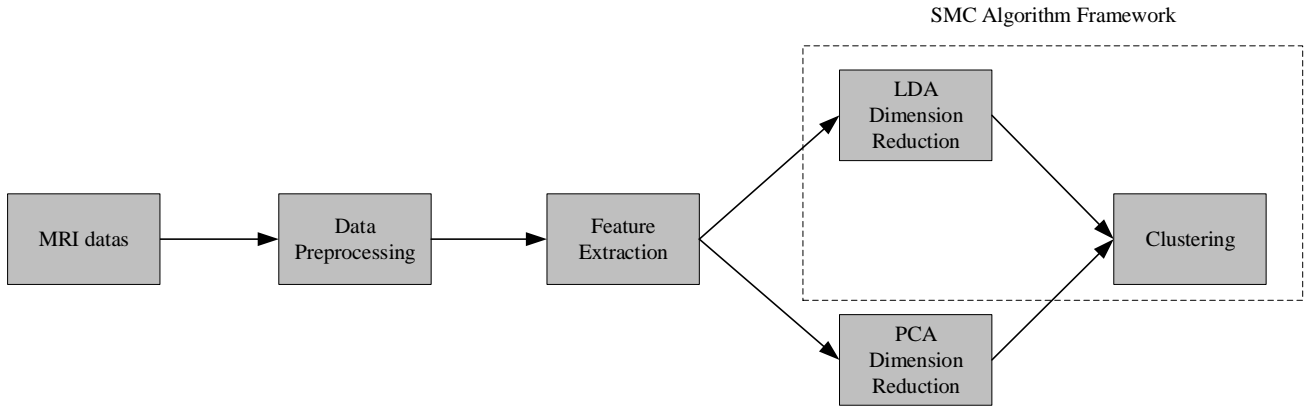

**Figure 6. The experimental design flowchart.**

## 3. Results

In this section, we firstly give the comparison of experiments results with different clustering methods and our proposed SMC. And then we demonstrate the effectiveness of SMC for screening PD. At last, all the different feature views are analyzed only by SMC for the screening and identification of PD.

*3.1. The compared results*

In this subsection, to verify the effectiveness of semi-supervised technique, we evaluate the clustering results using SMC with those with traditional **U**nsupervised **C**lustering by **P**CA (UCP) on 6 single-view clustering methods, including GMM, K-Means, K-Medoids, AC, Birch, SC and 1 multi-view clustering algorithm, namely RMKMC. Because there are 7 feature views i.e. Contrast, Homogeneity, Energy, Correlation, Sigma, Skew and Kurtosis, of which dimensionality reduction are processed by our proposed SMC and traditional PCA respectively, we implement each baseline algorithm on each view data and the compared results of SMC and PCA on different clustering methods are presented in Figure 7. Meanwhile, to show that our proposed model is good at feature extraction and enhancing the performance of clustering, we also utilize the SMC and UCP on each single feature view, and the compared result of different views are displayed in Figure 8.

*Mathematical Biosciences and Engineering*  Volume x, Issue x, 1-X Page.

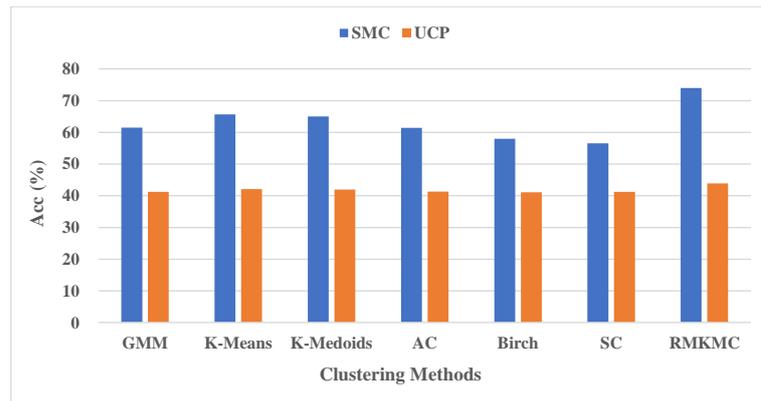

(a)

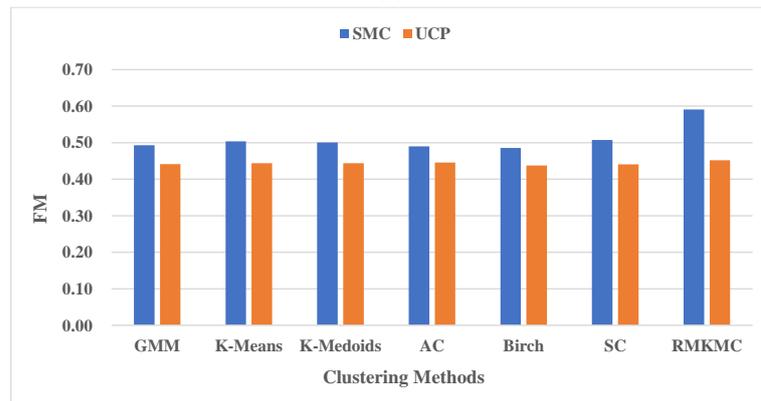

(b)

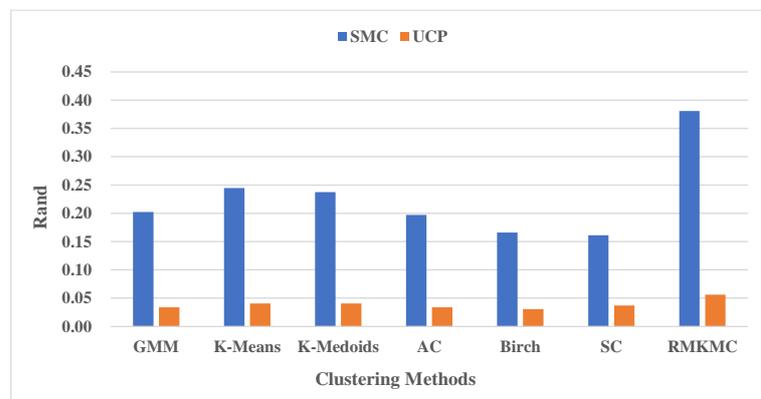

(c)

**Figure 7. The compared results of (a) Acc values, (b) FM values, (c) Rand values for different clustering methods by our SMC and UCP.**

Figure 7 shows the comparison results of different clustering algorithms after dimensionality reduction using our SMC framework and UCP. In this figure, we average the value of the clustering results of all the 7 views (i.e. Contrast, Homogeneity, Energy, Correlation, Sigma, Skew and Kurtosis) for the baselines, except for the multi-view clustering algorithm RMKMC. Compared with the single-view algorithms, it can be clearly noted that the SMC framework is superior to capture the significant features among MRIs than those which directly utilize UCP to do dimensionality reduction clustering according to the evaluation of ACC, FM and Rand values. As for the multi-view method, it is noted that our SMC has the better performance and it illustrates that SMC can exploit the LDA to train a more robust model and then improve the performance of clustering.

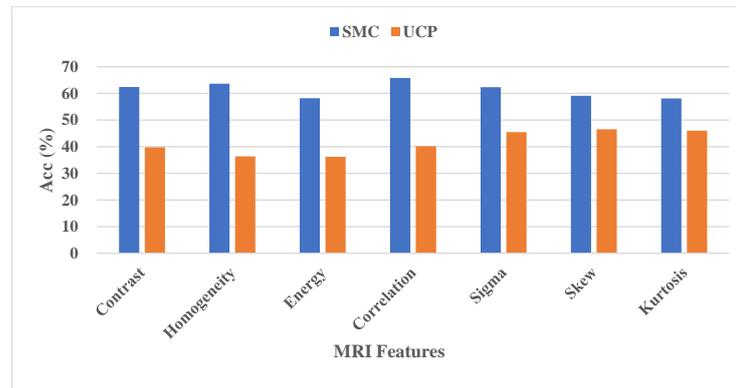

(a)

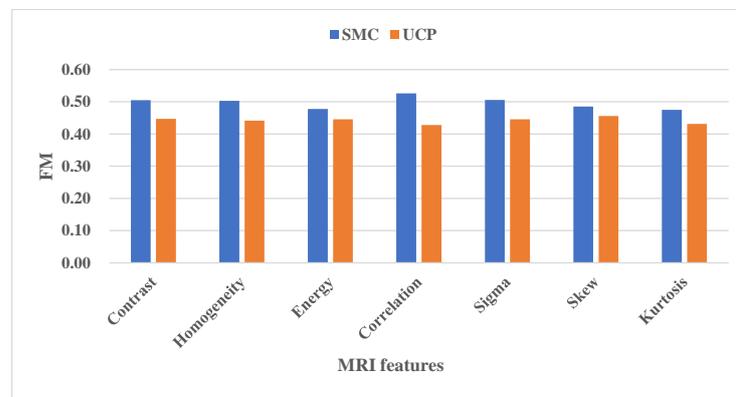

(b)

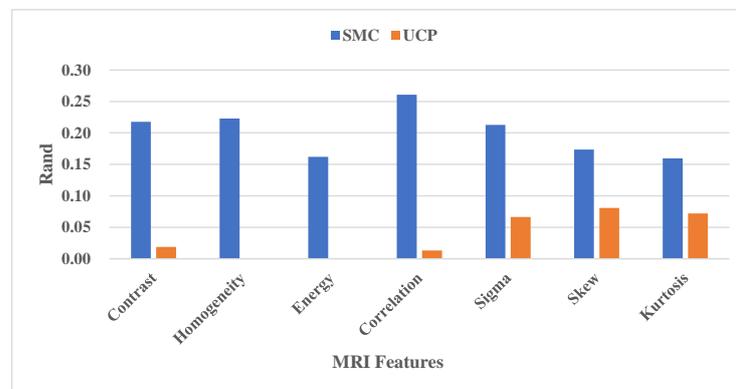

(c)

**Figure 8. The compared results of (a) acc values, (b) FM values, (c) Rand values for different MRI Features by our SMC and UCP.**

Figure 8 shows the comparison clustering results of different feature views after dimensionality reduction with our SMC and UCP. In this figure, the clustering values of each feature view are the average values of the GMM, K-means, K-means, AC, Birch, SC and RMKMC. It is obvious that the dimensionality reduction results from our proposed SMC framework is more appropriate than those from UCP on different evaluation metrics. Moreover, it is worth noting that our SMC is more stable than UCP when extracting effective information from different views.

## 3.2. *The discovery of efficient clustering methods and feature view for screening PD*

In order to find out the best clustering algorithm for screening PD and demonstrate the superiority of multi-view learning technology, we also compare the performance of different clustering methods with our proposed SMC. The detailed results are shown in the Table 2, 3, 4.

**Table 2. The Acc values of each clustering with SMC for each view** (%)

|  | GMM | K-Means | K-Medoids | AC | Birch | SC | Average | RMKMC |
|---|---|---|---|---|---|---|---|---|
| Contrast | 63.0±13.4 | **66.5±12.1** | 66.1±11.2 | 62.4±13.4 | 59.5±18.0 | 57.0±18.4 | 62.4±14.4 |  |
| Homogeney | 64.1±12.8 | **66.9±14.5** | 66.5±13.6 | 63.2±16.2 | 59.9±16.6 | 61.0±18.8 | 63.6±15.4 |  |
| Energy | 57.1±12.4 | 62.2±12.9 | **62.3±13.4** | 59.6±12.0 | 54.4±20.4 | 53.5±20.5 | 58.2±15.3 |  |
| Correlation | **66.6±13.7** | **69.8±17.1** | **69.3±14.6** | **64.4±13.8** | **60.7±16.0** | **64.1±20.6** | **65.8±16.0** |  |
| Sigma | 62.4±12.9 | **68.0±11.2** | 66.9±12.8 | 62.7±13.6 | 60.2±20.9 | 53.4±21.2 | 62.3±15.4 |  |
| Skew | 59.3±11.3 | **63.7±13.6** | 62.7±13.3 | 60.0±14.7 | 55.9±18.4 | 53.1±18.4 | 59.1±15.0 |  |
| Kurtosis | 58.2±14.9 | **62.5±14.0** | 61.3±15.0 | 57.6±19.1 | 55.0±18.4 | 53.9±20.3 | 58.1±17.0 |  |
| Average | 61.5±13.1 | **65.7±13.6** | 65.0±13.4 | 61.4±14.7 | 57.9±18.4 | 56.6±19.7 |  | **74 ± 8.9** |

Table 2 shows the Acc results of different clustering algorithms with different feature views after using SMC algorithm framework. Obviously, the multi-view clustering method is superior to other baselines, at around **74±8.9,** which illustrates that multi-view algorithms are capable of capturing sufficient features. As for the single view clustering methods, the algorithm with best average effect of the 7 feature views is K-Means in most cases, which is **65.7±13.6**. The best average effect of six single view clustering algorithms is the **Correlation** feature view data set, which is **65.8±16.0**. In the single view clustering methods, the all the highest values of accuracy are bold. Also, all the single view clustering methods obtain the highest Acc value in the **Correlation** feature view data set. From different feature views, the best single view clustering Acc of the seven characteristics are also bold, respectively.

**Table 3. The FM values of each single view clustering with SMC for each view**

|  | GMM | K-Means | K-Medoids | AC | Birch | SC | Average | RMKMC |
|---|---|---|---|---|---|---|---|---|
| Contrast | 0.50 ±0.10 | 0.51 ±0.09 | 0.51 ±0.10 | 0.50 ±0.10 | 0.49 ±0.10 | 0.51 ±0.10 | 0.50 ±0.10 |  |
| Homogeney | 0.50 ±0.09 | 0.51 ±0.14 | 0.51 ±0.14 | 0.50 ±0.10 | 0.49 ±0.07 | 0.51 ±0.10 | 0.50 ±0.11 |  |
| Energy | 0.47 ±0.07 | 0.48 ±0.06 | 0.48 ±0.06 | 0.47 ±0.07 | 0.48 ±0.09 | 0.49 ±0.08 | 0.48 ±0.07 |  |
| Correlation | **0.53 ±0.10** | **0.54 ±0.10** | **0.54 ±0.10** | **0.51 ±0.09** | **0.51 ±0.10** | **0.53 ±0.11** | **0.53 ±0.10** |  |
| Sigma | 0.50 ±0.08 | 0.52 ±0.12 | 0.51 ±0.10 | 0.49 ±0.09 | 0.49 ±0.09 | 0.52 ±0.08 | 0.51 ±0.09 |  |
| Skew | 0.48 ±0.09 | 0.49 ±0.10 | 0.49 ±0.09 | 0.48 ±0.09 | 0.48 ±0.08 | 0.50 ±0.09 | 0.49 ±0.09 |  |
| Kurtosis | 0.47 ±0.08 | 0.48 ±0.08 | 0.47 ±0.09 | 0.47 ±0.09 | 0.47 ±0.09 | 0.49 ±0.08 | 0.48 ±0.09 |  |
| Average | 0.49 ±0.09 | 0.50 ±0.10 | 0.50 ±0.10 | 0.49 ±0.09 | 0.49 ±0.09 | **0.51 ±0.09** |  | **0.59±0.11** |

Table 3 shows the FM results of different clustering algorithms and different feature views after using SMC algorithm framework. Similarly, the best FM value is **0.59±0.11** combining different feature view by RMKMC. And all the baselines achieve best performance on the Correlation feature view, at about **0.53 ±0.10**. It shows that Correlation view contains more effective and important information than others. Out of the six single view clustering methods, the average effect of SC clustering method is superior to other baselines.

**Table 4. The Rand values of each single view clustering with SMC for each view**

| Rand | GMM | K-Means | K-Medoids | AC | Birch | SC | Average | RMKMC |
|---|---|---|---|---|---|---|---|---|
| Contrast | 0.22±0.17 | 0.26±0.15 | 0.26±0.14 | 0.21±0.16 | 0.19±0.19 | 0.17±0.18 | 0.22±0.17 | |
| Homogeneiy | 0.23±0.17 | 0.26±0.23 | 0.25±0.22 | 0.22±0.18 | 0.18±0.16 | 0.20±0.20 | 0.22±0.19 | |
| Energy | 0.15±0.13 | 0.20±0.13 | 0.20±0.13 | 0.17±0.13 | 0.14±0.17 | 0.12±0.16 | 0.16±0.14 | |
| Correlation | **0.27±0.17** | **0.31±0.18** | **0.30±0.16** | **0.24±0.16** | **0.20±0.17** | **0.25±0.24** | **0.26±0.18** | |
| Sigma | 0.21±0.14 | 0.27±0.18 | 0.26±0.16 | 0.21±0.15 | 0.19±0.19 | 0.14±0.19 | 0.21±0.17 | |
| Skew | 0.17±0.13 | 0.22±0.17 | 0.21±0.16 | 0.18±0.15 | 0.14±0.16 | 0.12±0.14 | 0.17±0.15 | |
| Kurtosis | 0.16±0.16 | 0.20±0.16 | 0.19±0.17 | 0.16±0.18 | 0.13±0.18 | 0.13±0.15 | 0.16±0.17 | |
| Average | 0.20±0.15 | **0.25±0.17** | 0.24±0.16 | 0.20±0.16 | 0.17±0.17 | 0.16±0.18 | | **0.38±0.17** |

Table 4 shows the Rand results of different clustering algorithms on different feature views after using SMC algorithm framework. Also, the multi view clustering algorithm RMKMC combined with different features is still the best, of which Rand value is **0.38±0.17**. The best Rand average value of six single view clustering methods is from the **Correlation** feature view, which is **0.26±0.18**. Overall, K-Means has the better performance of Rand in out of the 6 single-view algorithms, which is **0.25±0.17**.

From the Table 2, 3 and 4, it demonstrates that the Correlation feature view is more essential and contains more effective information than other six feature views for screening PD under our proposed SMC. Simultaneously, the K-Means clustering method achieves the best overall performance not only in Table 2 but also in Table 3 and 4 compared with all the single view clustering models. Moreover, due to the ability of integrating different features from multiple views, RMKMC is more suitable to mine vital features from MRI data and to enhance the performance of clustering. Thus, the SMC framework based on RMKMC is capable of screening PD and assist in auxiliary diagnosis.

## 4. Discussion

With the development of artificial intelligence technology, the medical industry is paying more and more attention to use relevant methods to help prevent and diagnose diseases. At present, the global aging trend is relatively obvious, and PD is easy and common to occur in the elderly. Therefore, this study mainly uses machine learning technology to screen for PD through the human brain MRI data sets, which has the beneficial effect of prevention and auxiliary diagnosis. The experimental data set in the work is derived from an open data research platform i.e. PPMI.

In this paper, we mainly proposed a novel semi-supervised multi-view learning architecture with clustering (SMC) for screening PD. The SMC is a clustering learning framework based on LDA, the cross-validation method and RMKMC method. However, our SMC can also integrate the single view clustering methods i.e. GMM, K-Means, K-Medoids, AC, Birch and SC. In the experiments, we firstly preprocess the three-class data sets including non-patients, prodromal PD and confirmed PD by obtaining gray value images, removing noise points etc. And then the 7x7 sliding window method is used for extracting the 7 different feature data views containing Contrast, Homogeneity, Energy, Correlation, Sigma, Skew and Kurtosis. Finally, all the data sets are implemented into dimensionality reduction by SMC model and traditional PCA respectively, and compared by different clustering methods. It demonstrates that our proposed SMC framework outperforms other traditional model according to the comparison and analysis of the experimental results.

In the reality of examination and diagnosis for PD, there are only partially labeled brain MRI images. Thus, the excellent experimental results of SMC are very meaningful to provide the semi-supervised machine learning technology to assist in the diagnosis of PD with some data labels. In addition, this technology could be used in the physical examination of the elderly for screening latent PD, providing an early diagnosis and reducing the incidence of PD.

## 5. Conclusion

It is influential to use computer-aided diagnosis technologies with brain MRI data (or other types of medical data) to assist physicians to screen PD for making an exact diagnosis and treatment in modern medical industry. In this paper, we proposed a novel semi-supervised multi-view clustering architecture (denoted as SMC). SMC reconciles the key method of LDA, the cross-validation idea, semi-supervised, multi-view learning and clustering technologies. The original data sets from the PPMI database were first preprocessed by the optimization techniques mentioned above. And, the 7 view data sets of different features were extracted by the 7x7 sliding window method from the preprocessed data sets. Then, SMC learns a new representation of dimensionality reduction among multiple data views by training with LDA method. After training, SMC can effectively screen non-patients, prodromal PD and confirmed PD with different clustering models. Extensive experimental results demonstrated that the proposed model outperforms the traditional non-supervised clustering methods.

As we only evaluate our proposed SMC using MRI data in this work, we will work on collecting text views related to other fields in the future, explore the relationship between different views and further perform our model on these data sets. In addition, we will improve SMC to acquire higher clustering results for screening PD by considering different importance of distinct feature views.

## Acknowledgments

We would like to thank the Parkinson's Progression Markers Initiative (PPMI) for the datasets used in our experiments. This work is supported by the National Science Foundation of China (Nos. 61976247 and 61572407), the National Key Technology Research and Development Program (No. 2015BAH19F02).

## Conflict of Interest

All authors declared that we have no conflicts of interest to this work.

## References

1. C. W. Tsai, R. T. Tsai, S. P. Liu, et al, Neuroprotective effects of betulin in pharmacological and transgenic Caenorhabditis elegans models of parkinsons disease, *Cell Transplantation*, **26** (2017), 1903-1918.
2. R. E. Burke and K. O'Malley, Axon degeneration in parkinson's disease, *Experimental Neurology*, **246** (2013), 72-83.
3. C. P. Weingarten, M. H. Sundman, P. Hickey, et al, Neuroimaging of parkinson's disease: Expanding views, *Neuroscience and Biobehavioral Reviews*, **59** (2015), 16-52.
4. Y. Kim, S. M. Cheon, C. Youm, et al, Depression and posture in patients with parkinsons disease, *Gait & posture*, **61** (2018), 81-85.
5. R. Martínez-Fernández, R. Rodríguez-Rojas, M. del Álamo, et al, Focused ultrasound subthalamotomy in patients with asymmetric Parkinson's disease: a pilot study, *The Lancet Neurology*, **17** (2018), 54-63.
6. D. Frosini, M. Cosottini, D. Volterrani, et al, Neuroimaging in parkinson's disease: Focus on substantia nigra and nigro-striatal projection, *Current Opinion in Neurology*, **30** (2017), 416-426.
7. K. Marek, D. Jennings, S. Lasch, et al, The Parkinson progression marker initiative (PPMI), *Progress in Neurobiology*, **95** (2011), 629–635.
8. J. Shi, Z. Xue, Y. Dai, et al, Cascaded multi-column RVFL+ classifier for single-modal neuroimaging-based diagnosis of Parkinson's disease, *IEEE Transactions on Biomedical Engineering*, 66(2018), 2362-2371.
9. B. Peng, S. Wang, Z. Zhou, et al, A multilevel-roi-features-based machine learning method for detection of morphometric biomarkers in parkinsons disease, *Neuroscience Letters*, **651** (2017), 88–94.
10. R. Prashanth, S. D. Roy, P. K. Mandal, et al, High-accuracy classification of parkinson's disease through shape analysis and surface fitting in $^{123}$I-Ioflupane SPECT imaging, *IEEE Journal of Biomedical and Health Informatics*, **21** (2016), 794–802.
11. F. P. Oliveira and M. Castelo-Branco, Computer-aided diagnosis of Parkinson's disease based on [$^{123}$I] FP-CIT SPECT binding potential images, using the voxels-as-features approach and support vector machines, *Journal of Neural Engineering*, **12** (2015), 026008.
12. G. Garraux, C. Phillips, J. Schrouff, et al, Multiclass classification of FDG PET scans for the distinction between Parkinson's disease and atypical parkinsonian syndromes, *NeuroImage: Clinical*, **2** (2013), 883–893.
13. D. Long, J. Wang, M. Xuan, et al, Automatic classification of early Parkinson's disease with multi-modal MR imaging, *Plos One*, **7** (2012), e47714.
14. A. Abos, H. C. Baggio, B. Segura, et al, Discriminating cognitive status in Parkinson's disease through functional connectomics and machine learning, *Scientific Reports*, **7** (2017), 45347.
15. H. Lei, Z. Huang, F. Zhou, et al, Parkinson's disease diagnosis via joint learning from multiple modalities and relations, *IEEE Journal of Biomedical and Health Informatics*, 23(2019), 1437-1449.
16. E. Adeli, F. Shi, L. An, et al, Joint feature-sample selection and robust diagnosis of Parkinson's disease from MRI data, *NeuroImage*, **141** (2016), 206–219.
17. E. Adeli, G. Wu, B. Saghafi, et al, Kernel-based joint feature selection and max-margin classification for early diagnosis of parkinsons disease, *Scientific Reports*, **7** (2017), 41069.
18. X. Cai, F. Nie and H. Huang, Multi-view k-means clustering on big data, *Proceedings of the 23rd International Joint Conference on Artificial Intelligence*, Beijing,China, August 3-9, pp. 2598-2604.

19. R. Kohavi, A study of cross-validation and bootstrap for accuracy estimation and model selection, *Proceedings of the 14th International Joint conference on Artificial Intelligence, Montreal, Canada, August 19-20,* pp. 1137-1143
20. S. Balakrishnama and A. Ganapathiraju, Linear discriminant analysis-a brief tutorial, *Institute for Signal and information Processing*, **18** (1998), 1-8.
21. F. Samaria and F. Fallside, Face identification and feature extraction using hidden markov models, *Image Processing: Theory and Applications*, 4(1993), 295-298.
22. N. Vlassis and A. Likas, A greedy EM algorithm for Gaussian mixture learning, *Neural Processing Letters*, **15** (2002), 77-87.
23. D. Steinley, K-means clustering: a half-century synthesis, *British Journal of Mathematical and Statistical Psychology*, **59** (2006), 1-34.
24. H. S. Park and C. H. Jun, A simple and fast algorithm for K-medoids clustering, *Expert Systems with Applications*, **36** (2009), 3336-3341.
25. T. Kurita, An efficient agglomerative clustering algorithm using a heap, *Pattern Recognition*, **24** (1991), 205-209.
26. T. Zhang, R. Ramakrishnan and M. Livny, BIRCH: an efficient data clustering method for very large databases, *ACM Sigmod Record*, **25** (1996), 103-114.
27. U. Von Luxburg, A tutorial on spectral clustering, *Statistics and Computing*, **17** (2007), 395-416.
28. N. Wang, H. Yang, C. Li, et al, Using 'swallow-tail'sign and putaminal hypointensity as biomarkers to distinguish multiple system atrophy from idiopathic Parkinson's disease: A susceptibility-weighted imaging study, *European Radiology*, **27** (2017), 3174-3180.
29. K. Machhale, H. B. Nandpuru, V Kapuret, et al, MRI brain cancer classification using hybrid classifier (SVM-KNN), *2015 International Conference on Industrial Instrumentation and Control (ICIC)*, *Pune, India, May 28-30,* pp.60-65
30. P. Refaeilzadeh, L. Tang and H. Liu, "Cross-validation," in Encyclopedia of Data Base Systems. Cross-validation, *Encyclopedia of Database Systems*, Springer, New York, NY, USA, 2009.
31. M. Kirby, *Geometric data analysis: an empirical approach to dimensionality reduction and the study of patterns*, John Wiley & Sons, Inc., New York, NY, USA, 2001.
32. K. Honda and H. Ichihashi, Fuzzy local independent component analysis with external criteria and its application to knowledge discovery in databases, *International Journal of Approximate Reasoning*, **42** (2006), 159-173.
33. D. Donoho and V. Stodden, When does non-negative matrix factorization give a correct decomposition into parts?, *Proceedings of the 16$^{th}$ International Conference on Neural Information Processing Systems*, *Vancouver, BC, Canada, December* 8-13, 2003. pp.1141-1148.
34. A. Tharwat, Principal component analysis-a tutorial, *International Journal of Applied Pattern Recognition*, **3** (2016), 197-240.
35. T. Hastie and R. Tibshirani, Discriminant analysis by Gaussian mixtures, *Journal of the Royal Statistical Society: Series B (Methodological)*, **58** (1996), 155-176.
36. G. E. Hinton and R. R. Salakhutdinov, Reducing the dimensionality of data with neural networks, *Science*, **313** (2006), 504-507.
37. S. Mika, et al, Fisher discriminant analysis with kernels, *Proceedings of the 1999 IEEE Signal Processing Society Workshop, Madison, WI, USA, August 25-25. 1999,* pp. 41-48.
38. Y. Yang, H. Wang, Multi-view clustering: a survey, *Big Data Mining and Analytics*, **1** (2018), 83-107.

39. U. Maulik and S. Bandyopadhyay, Performance evaluation of some clustering algorithms and validity indices, *IEEE Transactions on Pattern Analysis and Machine Intelligence*, **24** (2002), 1650-1654.

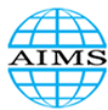